\documentclass[aps,prb,twocolumn,showpacs,amsmath,amssymb,floatfix]{revtex4}
\usepackage{graphicx}
\usepackage{epsfig}
\usepackage{bm}

\def\l{\langle\!\langle}
\def\r{\rangle\!\rangle}

\begin{document}

\title{Universal non-equilibrium states at the fractional quantum Hall edge}

\author{Ivan P. Levkivskyi}
\affiliation{
Institute for Theoretical Physics, ETH Zurich, CH-8093 Zurich, Switzerland
\\ 
Department of Physics, Harvard University, 17 Oxford Street, Cambridge MA 02138}
\date{\today}

\begin{abstract}
Integrability of electron dynamics in one dimension is manifested by the 
non-equilibrium stationary states. They emerge near a point contact coupling 
two quantum Hall edges with different chemical potentials. I use the 
non-equilibrium bosonization technique to show that the effective 
temperature of such states at the fractional 
quantum Hall edges has a universal linear dependence on the current through the
contact. In contrast, the temperature at eventual equilibrium scales as the 
square root of the power dissipating at the point contact.
I propose to use this distinction to detect these 
intriguing non-equilibrium states. 
\end{abstract}


\maketitle

\section{Introduction}

The two-dimensional electron gas (2DEG) in the regime of quantum Hall (QH) 
effect is a fascinating 
example of a system exhibiting rich emergent physics \cite{Klitz, QH-review}. 
In this regime, observed in the strong magnetic field, electrons in the bulk of 2DEG 
form an incompressible liquid with 
elementary excitations having fractional charge and fractional 
statistics \cite{Laugh, QH-review}. At the same time, chiral 
one-dimensional edge states \cite{Halp-edge} are present at the boundary of the 2DEG. 
Chiral nature of these collective states, that could be viewed as incompressible deformations 
of the electron liquid, makes them similar to optical beams. Such similarity has 
led to the emergence of a new sub-field of condensed matter physics, 
electron optics \cite{el-optics-new}.

The QH edge states can be described in the framework of the low-energy effective
theory approach \cite{eff-theory}, where the effective Hamiltonian is constructed 
from the general considerations of 
locality and gauge invariance. Such approach has a great success in describing 
recent electron optics experiments in the regime of integer QH effect, 
where an integer number $\nu$ of 
Landau levels is filled with electrons \cite{el-optics,Pierre}.
In particular, it has been fruitfully applied to strongly out-of-equilibrium 
situations, which take place in experiments on dephasing of the edge states 
\cite{heiblum-lobes, lobes} and on the energy relaxation at the QH edge 
\cite{Pierre,twoqpcs, prokudina}. This progress is due to the recent development of 
the technique of non-equilibrium bosonization \cite{our-phas}, which allows to treat 
non-perturbatively Coulomb interactions of one dimensional electrons
far from equilibrium.

\begin{figure}[t]
\begin{center}
\epsfxsize=8cm
\epsfbox{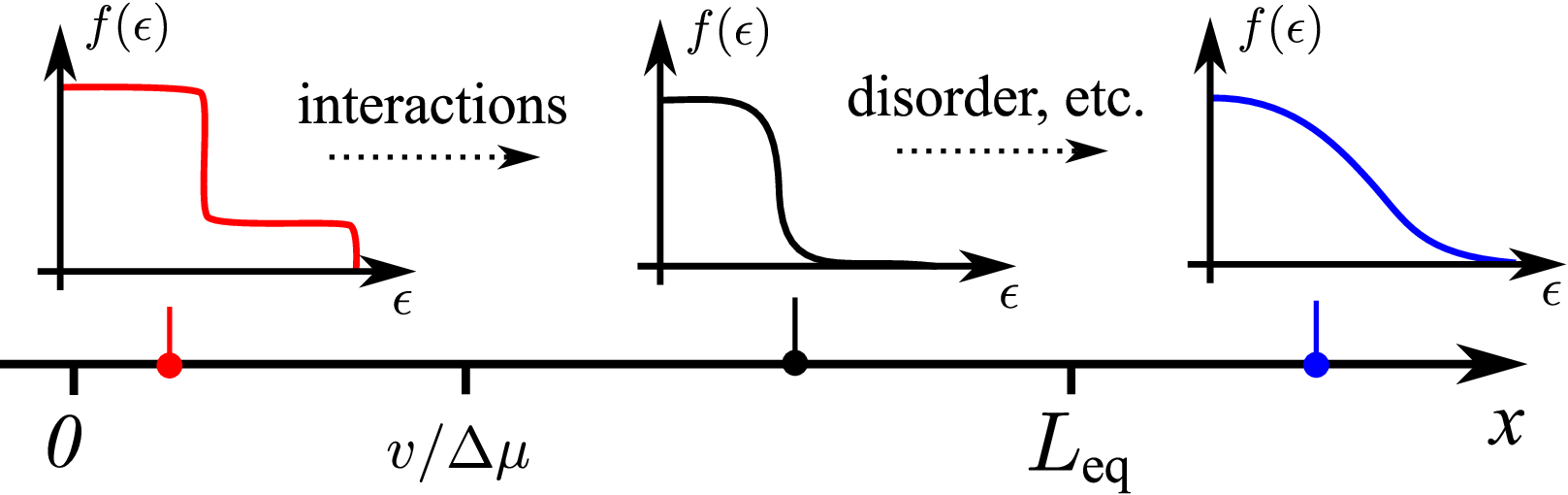}
\end{center}
\caption{(Color on-line) Illustration of the evolution of the energy 
distribution function at an integer QH edge. A biased QPC creates the 
non-equilibrium distribution function, sketched by the red curve. Due to 
interactions at the edge, it evolves to the intermediate distribution (black curve) 
at distances proportional to the velocity of edge excitations $v$. Finally, the 
distribution function equilibrates to the Fermi distribution (blue curve) at 
distances $L_{\rm eq}\gg v/\Delta\mu$, where additional mechanisms start to act 
\cite{our-rel}.}
\label{intermed}
\end{figure}

Non-equilibrium systems exhibit reach physical phenomena such as, e.g.,  
{\em pre-thermalization} of cold atomic gases, driven out of equilibrium by a quench 
\cite{cold-atoms}. One of the most common ways of preparing a non-equilibrium 
state is coupling two equilibrium systems with different parameters. For instance, 
non-equilibrium QH edge states are typically created by bringing two edges with 
chemical potential difference $\Delta\mu$ close to each other, 
so that electrons can tunnel across the so formed
quantum point contact (QPC). It has been found in Ref.\ [\onlinecite{our-rel}]
that there are two stages in the process of equilibration of such state at 
an integer QH edge. First, Coulomb interactions lead to relaxation to an 
intermediate non-equilibrium state, then additional energy exchange mechanisms, 
such as, e.g., disorder, equilibrate electrons to an actual Fermi
distribution. This two-stage behaviour is manifested via an 
intermediate asymptotic in the energy
distribution function of electrons downstream of a biased QPC, 
see Fig.\ \ref{intermed}. Note, that this effect is a consequence of the 
integrability of dynamics of interacting one-dimensional electrons.

\begin{figure}[t]
\begin{center}
\epsfxsize=8.5cm
\epsfbox{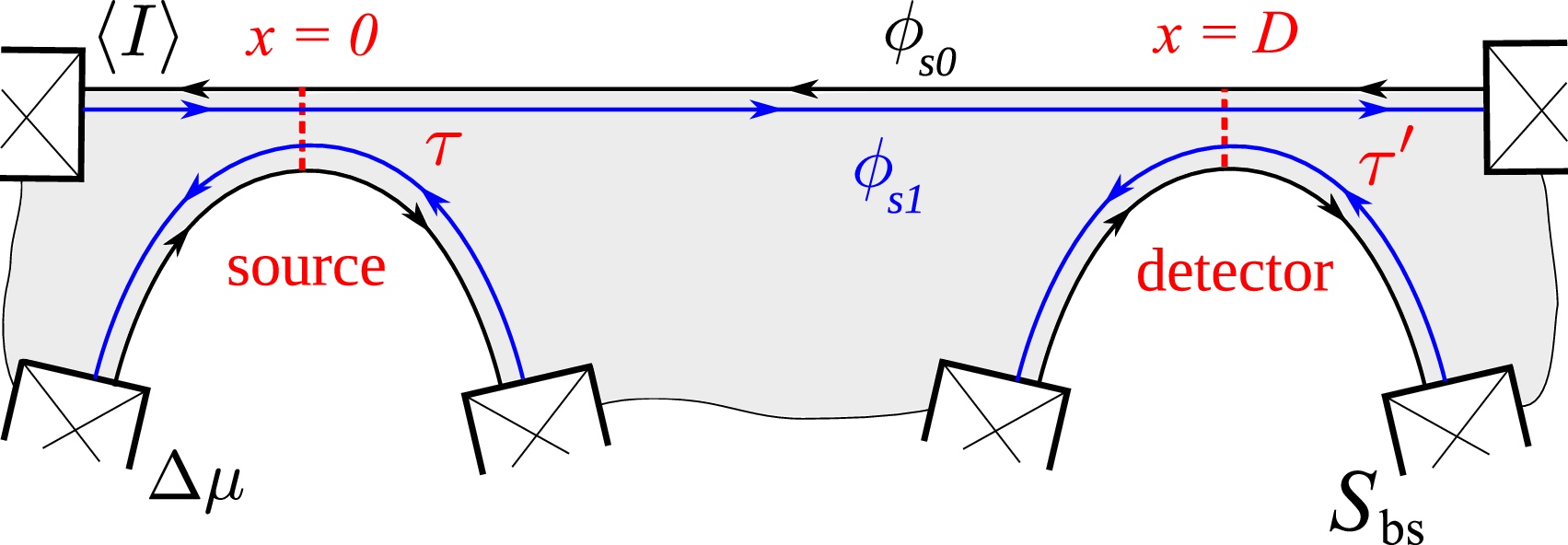}
\end{center}
\caption{(Color on-line) Experimental setup that can be used to detect the 
non-equilibrium stationary states at the QH edge. The 2DEG in the regime of 
fractional QH effect is shown by gray shadow. For a particular situation with 
two edge modes of opposite chiralities the left- and right-moving edge modes 
are shown by black and blue lines.
The left, upper, and right edges in the setup are labelled by indexes $s = L, U, R$.
The chemical potential of left edge is $\mu_L = \Delta\mu$, while 
$\mu_U = \mu_R = 0$. Tunnelling coupling between edge states are shown by 
red dashed lines. The source QPC creates a non-equilibrium state at the upper edge, 
the effective temperature of which is measured via the backscattering noise 
$S_{\rm bs}$ at the second, detector, QPC.
Such setup has been successfully used in Ref.~[\onlinecite{twoqpcs}] to measure 
the backscattering noise in the integer QH regime. 
}
\label{conc}
\end{figure}

The intermediate distribution function has a power-law asymptotic in contrast to 
the exponential decay of a Fermi distribution \cite{our-rel}.
However, a detection of the non-equilibrium states by measuring directly the 
distribution function might be intricate at typical values of tunnelling amplitude 
at the QPC, since the non-equilibrium distribution could be quantitatively similar 
to the Fermi distribution. One can overcome this obstacle by measuring a quantity 
which has qualitatively different behaviour in this state and at equilibrium. 
One of such quantities is the noise power of weak backscattering currents $j_{\rm bs}$ 
measured at a second, detector QPC \cite{twoqpcs, neder-wrong, our-thermo, bernd}, 
see Fig.\ \ref{conc}:
\begin{equation}
S_{\rm bs}\equiv \int dt \langle j_{\rm bs}(t) j_{\rm bs}(0)\rangle.
\label{n-p}
\end{equation}
It has been studied in Ref.\ [\onlinecite{our-thermo}] in the case of integer QH 
effect, where the intermediate asymptotic manifests itself in a non-analytic 
behaviour $S_{\rm bs}\propto \Delta\mu|\tau|^2\log |\tau|$ at small values of the 
tunnelling amplitude $\tau$. In contrast, the noise power behaves as 
$S_{\rm bs}\propto\Delta\mu|\tau|$ if a true thermalization occurs.

In this article, I investigate the non-equilibrium states in case of the 
{\em fractional} QH effect. The physics of fractional QH edge is more rich, 
since the fundamental excitations are not electrons but anyons with fractional 
charge and statistics \cite{Laugh}. Indeed, I find that 
the phenomenon of intermediate stationary state takes place in this case as well, 
and its physics is similar to discussed in Ref.\ [\onlinecite{our-thermo}]: 
Interactions alone can not equilibrate QH edge states because of integrability of 
their dynamics, but manifestations of this phenomenon are more dramatic as described 
below. I study the behaviour of the effective temperature of edge states, measured 
by one of the following detectors: by a second QPC via the backscattering noise 
power as discussed above (see Fig.\ \ref{conc}) and by a quantum anti-dot (QD) 
\cite{Pierre, amir} via QD level broadening (see Sec.\ \ref{lbr}). 
Na\"ively, one could think that these setups could be described using the 
perturbation theory in tunnelling at contacts. However, it has been shown in 
Ref.\ [\onlinecite{kane-perturb}] that such perturbation theory is divergent 
and fails at zero temperature. Therefore, I resort to the non-equilibrium 
bosonization \cite{our-phas, bagrets} and show that it can be used for a 
{\em non-perturbative} treatment of the source QPC in the weak backscattering 
limit, where rare scattering events of quasi-particles between edges have 
Poissonian statistics.

Using such approach, I find the central result of this paper: universal linear 
scaling of the effective temperature with the injected current,
\begin{equation}
\Theta_{\rm eff} \propto \langle I\rangle.
\label{noise-res}
\end{equation}
This scaling does not depend on the filling factor $\nu$ and is not modified neither by 
interactions at the edge nor by interactions between 
the edge states and the detector.
Moreover, it is drastically different from scaling that one expects 
after eventual equilibration of the edge states.
Indeed, at equilibrium at the temperature $\Theta$, the energy flux at the QH edge 
reads $\pi^2 \Theta^2/12$, see Ref.\ [\onlinecite{our-rel}]. 
Thus, the temperature of the edge states after a QPC where power 
$P=\Delta\mu\langle I\rangle$ is dissipated behaves as:
\begin{equation}
\Theta\propto \sqrt{P}.
\end{equation}
Such dependence is qualitatively different from (\ref{noise-res}) in the situation 
of fractional QH effect, where current-voltage characteristics of the QPC are 
highly non-linear. In addition, I clarify the reason for the perturbative 
divergence of noise mentioned above, it is the behaviour of the noise power 
$S_{\rm bs}\propto |\tau|^{4\delta - 2}$, that is not-analytic for the fractional 
values of the quasi-particle correlation function scaling dimension $\delta$.

In my analysis I focus on the $\nu = 2/m$ series of fractional QH states which are 
now extensively studied experimentally \cite{amir,exper-2m,neutral-exp}. An 
interesting feature of these series is the presence of 
neutral upstream modes. Such modes have been predicted theoretically in 
Ref.\ [\onlinecite{neutral-theor}] and experimentally detected recently 
\cite{neutral-exp}. 
They are interesting for two reasons: First, there is no average detector current, 
if the detector is upstream of the injection QPC, which makes the measurement of 
the effective temperature more simple.  Second, there is an opportunity to use 
the two-QPC setup to distinguish effective models with different values of coupling 
of quasi-particle excitations to the neutral mode.

\section{Effective theory of the edge states}

It has been shown that general constraints of 
locality and gauge invariance allow one to construct an effective theory of the QH 
edge states at energies much lower than the Fermi energy. \cite{Juerg-class} 
Such theory, however, 
contains arbitrary parameters, that can only be fixed experimentally or from 
microscopic calculations. Different realizations of these parameters are 
referred to as different effective models. 
In the minimal (i.e.\ simplest possible) effective models, the QH edge states 
at filling factors $\nu = 2/m$ with odd $m$ are 
described by boson fields $\phi_{s\alpha}(x,t)$, where $s = L,U,R$ enumerates the 
three edges in the experimental setup, left, upper and right, and $\alpha = 0,1$ 
enumerates the edge channels at each edge, see Fig.\ \ref{conc}. 
These fields have canonical commutation relations:
\begin{equation}
[\partial_x\phi_{s\alpha}(x), \phi_{s'\beta}(y)]=2\pi i 
(-1)^\alpha\delta_{\alpha\beta}\delta_{ss'}\delta(x-y),
\label{comm}
\end{equation}
where a different sign for $\alpha = 1$ reflects the opposite chirality of 
the corresponding channel. The charge densities at each edge in the system of 
units where $e = \hbar = 1$ are given by
\begin{equation}
\rho_s(x)= \frac{\sqrt{\nu}}{2\pi}[\cosh\theta\cdot\partial_x\phi_{s0}
(x)+\sinh\theta\cdot\partial_x\phi_{s1}(x)],
\label{rho-def}
\end{equation}
where $\theta$ is the first parameter that could not be fixed at the level of 
effective theory and describes the strength of interactions at the edge 
\cite{footnote-int,our-frac}.

The Hamiltonian of the setup depicted in the upper panel of Fig.\ \ref{conc} 
contains several terms ${\cal H}={\cal H}_0 + (A + A' + {\rm h.c.})$, 
where the first term describes the bare chiral dynamics of the edge excitations:
\begin{equation}
{\cal H}_0 = \sum_{s\alpha}\frac{v_\alpha}{4\pi}\int dx 
[\partial_x\phi_{s\alpha}(x)]^2.
\label{ham-0}
\end{equation}
Other terms describe weak backscattering at the two QPCs:
\begin{equation}
A = \tau \sum_\sigma e^{i[\eta_{L\sigma}(0)-\eta_{U\sigma}(0)]},\,
A' = \tau' \sum_\sigma e^{i[\eta_{U\sigma}(D)-\eta_{R\sigma}(D)]},
\label{tun-amp}
\end{equation}
where $\sigma = \pm$ enumerates the two types of local quasi-particle excitations 
with minimal charge 
$e^* = 1/m$ present in the model \cite{our-frac}:
\begin{multline}
\eta_{s\sigma}(x) = \frac{1}{\sqrt{2}}\left[\Big(\frac{\cosh\theta}
{\sqrt{m}}+\sigma\frac{\sinh\theta}{\sqrt{n}}\Big)\phi_{s0}(x) \right.\\ +\left.
\Big(\frac{\sinh\theta}{\sqrt{m}}+\sigma\frac{\cosh\theta}{\sqrt{n}}\Big)
\phi_{s1}(x)\right],
\end{multline}
where the odd number $n$ is the second parameter that could not be fixed 
at the level of effective theory. 
The question of which particular value of $n$ corresponds to actual 
experimental conditions could be answered either using microscopic 
ab initio calculations or experimentally \cite{our-frac}. 
Below I show that the experimental setups discussed in this article are also 
good tools for such task.

The equilibrium correlation functions of quasi-particles at 
temperature $\Theta$ and chemical 
potential $\mu_s$ in the effective theory described by 
Eqs.\ (\ref{comm}) and (\ref{ham-0}) are given by
\begin{subequations}
\begin{eqnarray}
\label{cf-eq}
\langle  e^{i\eta_{s\sigma}(x,t)} e^{-i\eta_{s\sigma}(x,0)} \rangle \propto 
\frac{\Theta^{\delta_\sigma} \exp{i\mu_s t/e^*}}
{[\sin\pi\Theta(it+0)]^{\delta_\sigma}},\\
\delta_\sigma = \frac{1}{2mn}\Big[(m+n)\cosh2\theta + 
2\sigma\sqrt{mn}\sinh2\theta\Big].
\end{eqnarray}
\end{subequations}
Consequently, the injection current in the weak tunnelling limit, given by the 
Kubo formula $\langle I\rangle = 
e^*\int dt \langle[A^\dag(t),A(0)]\rangle$, at low temperatures reads:
\begin{equation}
\langle I\rangle \propto |\tau|^2\Delta\mu^{2\delta - 1}, \hspace{12pt} 
\delta = \min(\delta_+, \delta_-).
\label{curr}
\end{equation}
Note that the fact that scaling exponent $\delta$ depends 
on the mixing angle $\theta$ is an obstacle in the experimental identification
of the topological parameter $n$. This also requires me to find a signature of 
the non-equilibrium stationary state that does not require the knowledge of the 
interactions strength. I show that the effective temperature measured in the 
two proposed schemes can be cast in a form that is not sensitive 
to the scaling $\delta$ at all. 
Therefore, without loss of generality, I focus on the situation of strong 
interactions $\theta = 0$, where one of the modes is completely neutral, 
see Eq.\ (\ref{rho-def}). I consider this situation to be the most relevant 
in view of the recent observation of neutral edge modes \cite{neutral-exp}.

\section{Noise power}

The backscattering currents operator at the detector QPC 
is given by a commutator of the total charge at the $s  = R$ edge with 
the total Hamiltonian. 
The result is $j_{\rm bs} = ie^*(A'-A'^\dag)$, so that in the leading order 
in $\tau'$, the noise power (\ref{n-p}) of these currents is given by
$S_{\rm bs} = (e^*)^2\int dt \langle\{A'^\dag(t), A'(0)\}\rangle$.
Using Eq.\ (\ref{tun-amp}) I find that the noise power is determined by the 
correlation functions of the quasi-particle operators:
\begin{multline}
S_{\rm bs} =2(e^*)^2|\tau'|^2\sum_\sigma\int dt \langle  e^{-i\eta_{U\sigma}(D,t)} 
e^{i\eta_{U\sigma}(D,0)}\rangle \\ \times \langle  e^{i\eta_{R\sigma}(D,t)} 
e^{-i\eta_{R\sigma}(D,0)} \rangle.
\label{noise-int}
\end{multline}
At equilibrium, the correlation function is given by Eq.\ (\ref{cf-eq}) and the 
integral in Eq.\ (\ref{noise-int}) evaluates to
\begin{equation}
S_{\rm bs} \propto \Theta^{2\delta - 1},
\label{temp-eq}
\end{equation}
where the quasi-particle's scaling dimension at $\theta=0$ is $\delta =1/2m+1/2n$. 
Thus, I define the effective temperature as $\Theta_{\rm eff}\equiv 
S_{\rm bs}^{1/(2\delta-1)}$, i.e., as a value that such QPC thermometer would 
show while probing an arbitrary state. In realistic systems, scaling dimension 
$\delta$ could deviate significantly from the theoretically predicted values 
\cite{chang-orig, chang-review}, therefore one needs to calibrate the thermometer 
on equilibrium states with known temperatures and use the experimentally found 
value of $\delta$. Note, that I assume that $\tau' \ll \tau$, so that the noise 
generated by the detector QPC is small, i.e., that such thermometer perturbs 
the state weakly.

\begin{figure}[t]
\begin{center}
\epsfxsize=8.5cm
\epsfbox{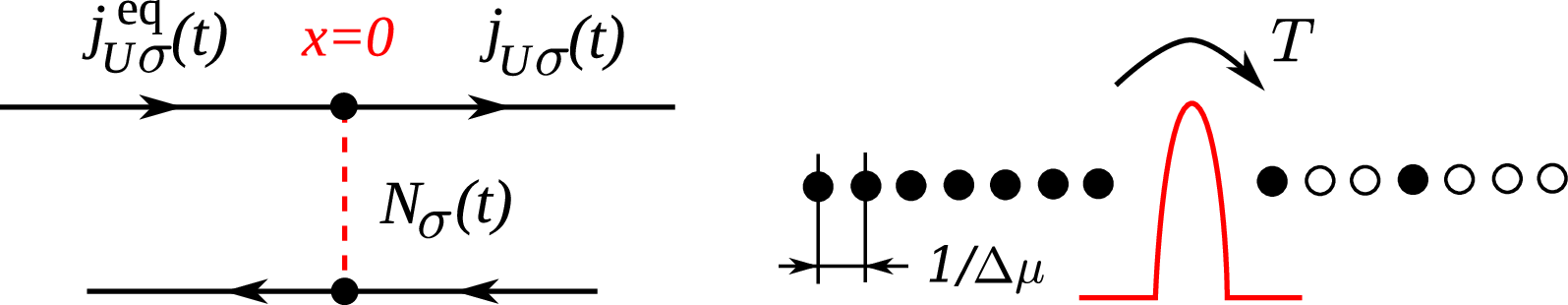}
\end{center}
\caption{(Color on-line) Illustrations for the non-equilibrium bosonization method 
and results. {\em Left panel:} Notations for the current operators at the upper 
edge near the injection QPC. The total currents $j_{U\sigma}(t)$ are represented 
as a sum of the equilibrium incoming contributions $j^{\rm eq}_{U\sigma}(t)$ 
and the tunnelling currents, the latter are expressed in terms of the number of 
quasi-particles operators $N_\sigma(t)$, which could be treated as classical 
stochastic quantities in the long-time limit. {\em Right panel:} Correlation 
time of quasi-particles at the edge is $\propto 1/\Delta\mu$, quasi-particles 
are independently backscattered with a small probability $T$ resulting in 
a Poissonian process with the average number of backscattered particles 
$\propto T\Delta\mu |t|$ on a time interval $t$.}
\label{simple}
\end{figure}

Assuming zero base temperature, the correlation function for $s=R$ is evaluated 
over the ground state $\langle  e^{i\eta_{R\sigma}(D,t)} 
e^{-i\eta_{R\sigma}(D,0)} \rangle \propto (it+0)^{-\delta}$. 
As discussed above, the perturbative theory is divergent and one needs to find 
a non-perturbative expression
for the correlation function at the upper, $s=U$, edge in a non-equilibrium 
state created by the backscattering processes at the source QPC. This could be done 
using the non-equilibrium bosonization technique proposed in 
Ref.\ [\onlinecite{our-phas}]. In this technique, 
the boson fields $\eta_{U\sigma}$ are expressed in terms of the backscattering 
currents $j_{U\sigma}$ at the source QPC by 
solving the equations of motion generated by Hamiltonian (\ref{ham-0}) 
with the boundary conditions at $x=0$:
\begin{equation}
\partial_t\eta_{U\sigma}(0,t) = 2\pi j_{U\sigma}(t).
\end{equation}
In the situation I consider, the edge dynamics is given by $\eta_{U\sigma}(D,t) = 
\eta^{\rm eq}_{U\sigma}(D,t) + 2\pi\sigma e^*\big[N_+(t-D/v_1) - N_-(t-D/v_1)\big]$, 
where $e^*N_{\sigma}(t)\equiv\int_{-\infty}^t\big[j_{U\sigma}(t') - 
j^{\rm eq}_{U\sigma}(t')\big]$ are the operators of the number of quasi-particles, 
see Fig.~\ref{simple}. 
At low energies the correlation function factorizes
\begin{equation}
\langle  e^{-i\eta_{U\sigma}(D,t)} e^{i\eta_{U\sigma}(D,0)}\rangle \propto 
\chi(t)(it+0)^{-\delta},
\label{corr-non-eq}
\end{equation}
where $\chi(t)$ is the purely non-equilibrium contribution expressed via the 
full counting statistics \cite{Levitov} of the quasi-particles' 
backscattering process:  
\begin{equation}\nonumber
\chi(t) = \prod_\sigma\chi_\sigma(2\pi\sigma/m, t), \, \chi_\sigma(\lambda, t) 
\equiv \langle e^{i\lambda N_{\sigma}(t)} e^{-i\lambda N_{\sigma}(0)}\rangle.
\end{equation}

Although in general 
the full counting statistics is a complex quantity, 
two important simplifications arise in the limit of weak backscattering: 
First, all the cumulants of quasi-particle numbers for the rare, Poissonian, 
process are equal to the average number of backscattered particles, i.e., 
for $p^{\rm th}$ cumulant one has $\l N_{\sigma}^p\r \equiv 
\partial_{i\lambda}^p\log\chi_\sigma(\lambda,t)|_{\lambda = 0} 
=\langle I\rangle|t|/2e^*$.  Note that factor 2 appears since we have two 
flavours of quasi-particles that both contribute to the current $\langle I\rangle$.
Thus, the non-equilibrium part of the correlation function is given by
\begin{equation}
\log\chi(t)\simeq \frac{\langle I\rangle|t|}{e^*}[\cos(2\pi/m) - 1], 
\hspace{10pt}\Delta\mu|t|\gg 1.
\label{chi-res}
\end{equation}
Second, the main contribution to integral (\ref{noise-int}) comes from large 
times $\sim 1/\langle I\rangle$, exactly where backscattering could be 
considered a classical Markovian process and result (\ref{chi-res}) is valid.

Indeed, although the correlation function (\ref{cf-eq}) for fractional 
quasi-particles has long-range character, the correlation time for tunnelling 
across QPC is finite and is of order $\sim 1/\Delta\mu$.
Mathematically, this stems from the oscillatory factors $\exp{i\Delta\mu t/e^*}$ 
that are present in correlation function Eq.\ (\ref{cf-eq}) and in all higher-order 
correlators, since such factors will cut-off arbitrary time integrals, see 
Ref.~[\onlinecite{Levitov}]. One can check this statement explicitly for lower order
cumulants. For instance, perturbative calculations for the noise of transmitted 
quasi-particle number at the injection QPC show that:
\begin{equation}
\langle \Delta N_\sigma^2(t)\rangle - \langle \Delta N_\sigma(t)\rangle^2 = 
\langle I\rangle|t|/2e^*, \hspace{12pt} \Delta\mu |t| \gg 1
\end{equation}
where $\Delta N_\sigma(t) \equiv N_\sigma(t)-N_\sigma(0)$, and it is natural to 
assume that the same behaviour holds for cumulants of arbitrary order $p$.
If the backscattering probability $T = \langle I\rangle/(\Delta\mu/2\pi m)$ is 
small, then the main contribution to the integral (\ref{noise-int}) comes from 
times that are much larger than the correlation time, where the correlation 
function has asymptotic
form given by Eq.~(\ref{chi-res}) that is insensitive to any complex details of 
backscattering at the fractional QH edge at short times.

Substituting expression (\ref{chi-res}) back in Eq.\ (\ref{corr-non-eq}) and 
then in Eq.\ (\ref{noise-int}) I arrive at the result $S_{\rm bs} \propto 
\langle I\rangle^{2\delta-1}$, so that comparing it to Eq.\ (\ref{temp-eq}) 
I indeed get the behaviour (\ref{noise-res}) independently of the scaling 
dimension $\delta$. Note that
the exponent $2\delta-1$ in the relation between the noise $S_{\rm bs}$ and current 
$\langle I\rangle$ is the same as in the relation between the current and 
the voltage bias $\Delta\mu$. This fact could be qualitatively understood from 
the following argument.
Both quantities (current and noise) 
are determined by the quasi-particle correlation functions. In both cases these 
correlation functions have the same scaling dictated by the orthogonality catastrophe, 
while the energy scale at which it is cut-off is different. It is given by 
bias or temperature at equilibrium (see Eq.\ (\ref{cf-eq})), and it is proportional 
to the current for the non-equilibrium correlation function 
(see Eq.\ (\ref{chi-res})).

An important remark is in order. Note that scaling of the backscattering noise 
power with the voltage bias at the injection QPC, 
$S_{\rm bs}\propto\Delta\mu^{(2\delta-1)^2}$, differs from scaling of the current 
(\ref{curr}) or of the equilibrium noise (\ref{temp-eq}). This means that it 
would be easier to experimentally distinguish different but close values of $\delta$ 
given by different effective models by measuring this non-equilibrium scaling. 
The experiment for measuring non-equilibrium scaling should be performed as in 
Ref.~[\onlinecite{twoqpcs}] (see Fig.~\ref{conc}) but for different distances 
between two QPCs, so that both non-equilibrium scaling at short distances and 
equilibrium scaling at long distances are detected.

One may think that the universal behaviour of the effective temperature found above 
is only specific of a particular QPC-based detection scheme. Therefore, I also show 
that a QD-based detection scheme leads to exactly the same result 
(see Sec.\ \ref{lbr}). Moreover, one can speculate that such behaviour is quite 
general for the situation of a weakly coupled thermometer. Indeed, such thermometer 
requires long time to measure the effective temperature and therefore it should be 
sensitive only to the long time asymptotic of the correlation functions at 
the edge, while the latter depend only on the average current in the universal 
manner, see Eq.\ (\ref{chi-res}).

\section{Level broadening}
\label{lbr}

\begin{figure}[t]
\begin{center}
\epsfxsize=8.5cm
\epsfbox{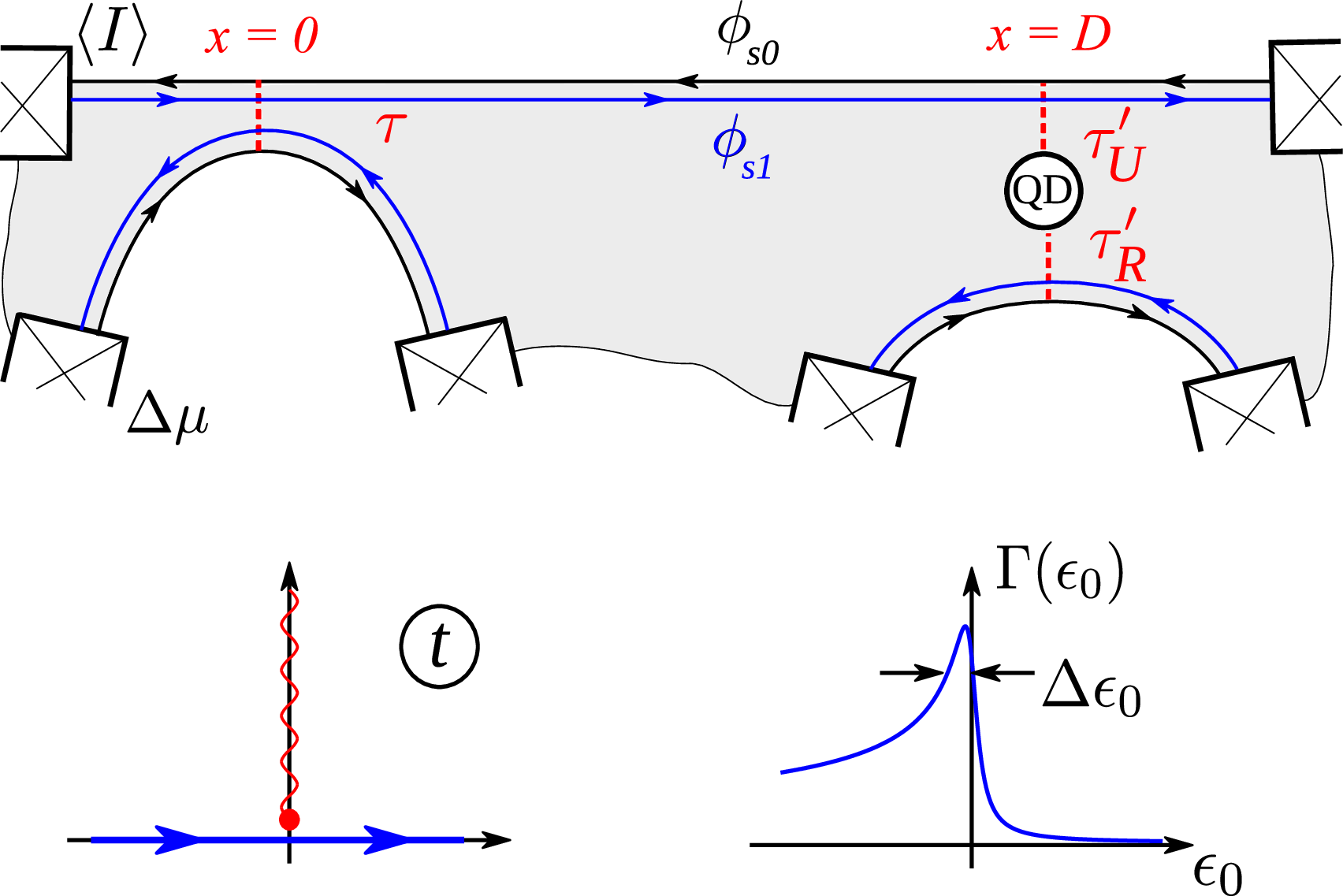}
\end{center}
\caption{(Color on-line) Details of an alternative measurement scheme for the 
effective temperature. {\em Upper panel:} While QPC is used as a source of noise, 
the detector QPC is now replaced with a QD coupled to the non-equilibrium upper 
edge. Effective temperature can be measured as level broadening of this QD. 
{\em Left panel:} 
The contour of integration in time integral of Eq. (\ref{gamma-int}) for the rate 
of tunnelling of quasi-particles to the QD is shown by blue line. The branch 
cut of the quasi-particle correlation functions goes along the imaginary axis 
and is shown by the red line. {\em Right panel:} The tunnelling rate $\Gamma$ 
of quasi-particles (\ref{gamma-res}) as a function of the QD energy level 
$\epsilon_0$ for the situation of strong interaction $g_U=g_R=1/2$ shows 
the level broadening $\Delta\epsilon_0$.}
\label{integr}
\end{figure}

To check the universality of the results found in the previous Section, I consider 
the second measurement scheme depicted in Fig.\ \ref{integr}. The effective 
temperature is measured as the width of the levels of a QD located 
upstream of the injecting QPC. For simplicity, I consider only one QD level, so that 
its Hamiltonian is ${\cal H}_{\rm QD}=\bar{\epsilon}_0\mathbf{d}^\dag \mathbf{d}$.
It has been shown experimentally \cite{heiblum}, that the Coulomb interactions of 
the edge states with 
this QD level could be strong and can not be neglected. The Hamiltonian of these 
interactions is 
\begin{equation}
{\cal H}_{\rm int}= \sum_{s = U,R}\int dx U_s(x)\rho_s(x)\mathbf{d}^\dag \mathbf{d},
\end{equation}
where $U_s(x)$ are screened Coulomb potentials. Such interaction term 
can be removed by an unitary transformation \cite{our-qd-Jura, glazman-deep-hole} 
${\cal H}\to S {\cal H} S^\dag$ with
\begin{equation}
S = \exp[iv_0^{-1}\sum_s\int dxU_s(x)\rho_{s}(x) {\bf d}^\dag{\bf d}]. 
\end{equation}
However, at the same time, assuming short range interactions 
$U_s(x) = U_s\delta(x-D)$, the tunnelling amplitudes after such transformation read
\begin{subequations}
\label{tun-dot}
\begin{eqnarray}
A'_U &=& \tau'_U \sum_\sigma \mathbf{d}^\dag \exp{[-ig_{R}\phi_{R0}(D)/\sqrt{2m}]}
\\ &\times & \exp{[i(1-g_U)\phi_{U0}(D)/\sqrt{2m}+i\sigma\phi_{U1}(D)/\sqrt{2n}]},
\nonumber
\end{eqnarray}
\begin{eqnarray}
A'_R &=& \tau'_R \sum_\sigma \mathbf{d}^\dag \exp{[-ig_{U}\phi_{U0}(D)/\sqrt{2m}]}
\\ &\times & \exp{[i(1-g_R)\phi_{R0}(D)/\sqrt{2m}+i\sigma\phi_{R1}(D)/\sqrt{2n}]}.
\nonumber
\end{eqnarray}
\end{subequations}

The dimensionless couplings $g_s \equiv U_s/\pi v_0$ take values between 0 and 1 
and have physical meaning of charges in units of $e^*$ accumulated at the 
corresponding channels due to interactions with the QD. As well, the energy 
$\bar{\epsilon}_0$ is renormalized \cite{our-qd-Jura} to $\epsilon_0 = 
\bar{\epsilon}_0 + \sum_s\int dx U^2_s(x)/v_0$.

There are two regimes in which level broadening is dominated by one of the 
two sources: 
First, quantum, source is tunnelling of the quasi-particles to and from the QD. 
Second, classical, source is the interaction with the non-equilibrium charge 
fluctuations which ``shake'' the QD level. I focus on the situation of classical 
level broadening, where 
the QD is almost totally incoherent and can be described by a master 
equation \cite{our-qd-Jura}.
The classical regime is realized if $G' \ll G \ll 1/2\pi$, where $G$ is the 
conductance of the injecting QPC, 
and $G'$ is the conductance of the detector. This is exactly the regime where 
thermometer perturbs the state weakly.
Level broadening in the classical regime can be found by studying the rate 
of tunnelling of quasi-particles from the upper channel to 
the QD, which is given by $\Gamma \equiv \int dt\langle A'^\dag_U(t)A'_U(0)\rangle$ 
in the leading order in $\tau'$. Using Eqs.\ (\ref{tun-dot}) 
and (\ref{chi-res}) I find that the tunnelling rate is given by
\begin{equation}
\Gamma \propto\int dt \chi(t)e^{i\epsilon_0t}(it+0)^{-\delta'},
\label{gamma-int}
\end{equation}
where $\delta' =[g_R^2+(1-g_U)^2]/2m+1/2n$, and the contour of integration goes as 
shown in Fig.\ \ref{integr}, middle panel.

Evaluating the integral, I find the tunnelling rate
\begin{equation}
\label{gamma-res}
\Gamma(\epsilon_0)\propto {\rm Re}[i(\epsilon_0 +i\Delta\epsilon_0)^{\delta'-1}],
\end{equation}
where $\Delta\epsilon_0 = \langle I\rangle[\cos (2\pi/m) -1]/e^*$.
A schematic plot of the rate (\ref{gamma-res}) for the case of strong interaction 
is shown in Fig.\ \ref{integr}, right panel.
It is not universal and depends on the interaction strength via couplings $g_s$. 
However, the level broadening $\Delta\epsilon_0$ itself has 
a universal linear dependence on the current so that the effective temperature is 
given by Eq.\ (\ref{noise-res}) again. This result serves as an additional argument 
in favour of the universality of relation (\ref{noise-res}) and shows that it is 
not a property of a particular detection scheme.

\section{Discussion of results}

Previous theoretical works \cite{our-rel, neder-wrong, our-thermo, bernd} showed 
that a non-equilibrium state created by coupling
two integer QH edges with different electro-chemical potentials at a QPC 
\cite{twoqpcs, amir} relaxes towards equilibrium via an intermediate asymptotic. 
This non-equilibrium stationary state is formed as a result of integrability of 
dynamics of one dimensional interacting electrons.
It was predicted that such state has a power-law energy distribution 
function \cite{our-rel} and generates backscattering noise with a non-analytic 
dependence on the transparency of the source QPC \cite{our-thermo}. 
Earlier attempts to investigate such physics in fractional QH states using 
perturbative techniques has led to divergences at low temperatures 
\cite{kane-perturb}.

In this paper I show that the non-equilibrium state created by coupling two 
{\em fractional} QH edges also relaxes through an intermediate stationary state 
that is qualitatively different from an equilibrium one. Namely, the 
non-equilibrium bosonization \cite{our-phas, bagrets} approach allows me to 
reduce the complex problem of non-perturbative treatment of fractional QH edge 
states to evaluating full counting statistics of a partitioned QPC, and 
ultimately to a simple Poissonian process, see Fig.\ \ref{simple}. 
Universal behaviour of Poissonian statistics leads to exponential decay
of the correlation function (\ref{chi-res}) similar to equilibrium one. 
That, in turn, leads to a possibility to express the the noise power
and the level broadening generated by intermediate state in terms of 
an effective temperature which has a universal linear dependence on the 
injected current.

Main physical consequence of these results is an experimental way to detect 
the pre-thermalization type phenomena \cite{cold-atoms} for fractional QH edge 
states, and essentially in arbitrary Luttinger liquid type system. 
Although recently observed bulk heat currents \cite{bulk-heat} in fractional 
QH systems could in principle compromise my results quantitatively, there 
is a qualitative difference between equilibrium and non-equilibrium scaling. 
In addition, the obtained results show that the measurement of scaling of the 
backscattering noise with the bias in such states can be used to efficiently 
distinguish effective models of QH edge states.

\begin{acknowledgements} I am grateful to I. Chernii, B.I. Halperin, B. Rosenow, 
A.O. Slobodeniuk, and E.V. Sukhorukov for several fruitful discussions and 
acknowledge the support of the Swiss NSF.
\end{acknowledgements}

\bibliographystyle{apsrev}

\begin{thebibliography}{99}
\bibitem{footnote-int}
Note that there is a particularly simple interpretation of this parameter in 
the situation of integer QH effect, where it plays the role of an angle of 
rotation between the bases of Landau levels and of edge plasmon eigenmodes.
\end{thebibliography}

\end{document}